\documentclass[fleqn,usenatbib]{mnras}

\usepackage{newtxtext,newtxmath}
\usepackage[T1]{fontenc}

\DeclareRobustCommand{\VAN}[3]{#2}
\let\VANthebibliography\thebibliography
\def\thebibliography{\DeclareRobustCommand{\VAN}[3]{##3}\VANthebibliography}

\usepackage{graphicx}	% Including figure files
\usepackage{amsmath}	% Advanced maths commands
\newcommand\cm{{\rm\thinspace cm}}

\newcommand\erg{{\rm\thinspace erg}}

\newcommand\K{{\rm\thinspace K}}
\newcommand\keV{{\rm\thinspace keV}}

\newcommand\kpc{{\rm\thinspace kpc}}
\newcommand\Lsun{\hbox{$\rm\thinspace L_{\odot}$}}

\newcommand\Msun{\hbox{$\rm\thinspace M_{\odot}$}}

\newcommand\s{{\rm\thinspace s}}
\newcommand\yr{{\rm\thinspace yr}}

\newcommand\chisq{\hbox{$\chi^2$}}

\newcommand\cmcu{\hbox{$\cm^3\,$}}

\newcommand\ergcmcups{\hbox{$\erg\cm^3\ps\,$}}

\newcommand\Msunpyr{\hbox{$\Msun\yr^{-1}\,$}}
\newcommand\pcm{\hbox{$\cm^{-3}\,$}}

\newcommand\pcmsq{\hbox{$\cm^{-2}\,$}}
\newcommand\pcmK{\hbox{$\cm^{-3}\K$}}

\newcommand\ps{\hbox{$\s^{-1}\,$}}
\newcommand\psqcm{\hbox{$\cm^{-2}\,$}}

\title[Hidden Cooling Flows]{Hidden Cooling Flows in  Clusters of Galaxies}

\author[A. C. Fabian et al.]{
A. C. Fabian,$^{1}$\thanks{E-mail: acf@ast.cam.ac.uk }, G.J. Ferland$^{2}$, J.S. Sanders$^{3}$, B.R. McNamara$^{4}$, C. Pinto$^{5}$ and S.A. Walker$^6$
\\
$^{1}$Institute of Astronomy, University of Cambridge, Madingley Road, Cambridge CB3 0HA, UK\\
$^{2} $Department of Physics, University of Kentucky, Lexington KY 40506, USA\\
$^{3} $Max-Planck-Institut fur extraterrestrische Physik, Giessenbachstrasse 1, 85748 Garching, Germany\\
$^{4} $Department of Physics and Astronomy, University of Waterloo, 200 University Avenue West, Waterloo, ON N2L 3G1, Canada\\
$^{5} $INAF-IASF Palermo, Via U. La Malfa 153, I-90146 Palermo, Italy\\
$^6$ Department of Physics and Astronomy, The University of Alabama in Huntsville, Huntsville, AL 35899, USA\\ }
\date{Accepted XXX. Received YYY; in original form ZZZ}

\pubyear{2022}

\begin{document}
\label{firstpage}
\pagerange{\pageref{firstpage}--\pageref{lastpage}}
\maketitle

\begin{abstract}
The radiative cooling time of the hot gas at the centres of cool cores in clusters of galaxies drops down to 10 million years and below. The observed mass cooling rate of such gas is very low, suggesting that AGN feedback is very tightly balanced or that the soft X-ray emission from cooling is somehow hidden from view. We use an intrinsic absorption model in which the cooling and coolest gas are closely interleaved to search for hidden cooling flows in the Centaurus, Perseus and A1835 clusters of galaxies. We find hidden mass cooling rates of between 10 to $500\Msunpyr$ as the cluster mass increases, with the absorbed emission emerging in the Far Infrared band. Good agreement is found between the hidden cooling rate and observed FIR luminosity in the Centaurus Cluster. The limits on the other two clusters allow for considerable hidden cooling. The implied total mass of cooled gas is much larger than the observed molecular masses. We discuss its fate including  possible further cooling and collapse into undetected very cold clouds, low mass stars and substellar objects,

\end{abstract}


%%%%%%%%%%%%%%%%%%%%%%%%%%%%%%%%%%%%%%%%%%%%%%%%%%

%%%%%%%%%%%%%%%%% BODY OF PAPER %%%%%%%%%%%%%%%%%%

\section{Introduction}

More than one half of clusters of galaxies have a cool core around their Brightest Cluster Galaxy or BCG. It ranges from 30 kpc to 100 kpc in radius depending on the mass and luminosity of the cluster, where the radius is defined  where the radiative cooling time of the gas is 7 Gyr, or similar fiducial value\footnote{7Gyr is the time since redshift one and may be considered to be the age of a low redshift cluster}. Inward of that radius, $r_{\rm c},$ the temperature of the hot IntraCluster Medium (ICM) drops and, to maintain hydrostatic equilibrium supporting the weight of the overlying gas, the density rises. Together the density $n$ and temperature $T$ combine, assuming dominant bremsstrahlung emission, to make the radiative cooling time $t_{\rm c} \propto T^{1/2}/n$ also drops inward, matching the expected appearance of a cooling flow. In the absence of any heating,  an inward flow of gas with a mass flow rate takes place 
\begin{equation}
\dot M = \frac{2}{5}\frac{L\mu m_{\rm H}}{kT},
\end{equation}
where $L$ is the luminosity of the region within $r_{\rm c}$. The rate may be halved when the gravitational work done on the flow is considered, since the initial temperature of the gas is usually close to the Virial temperature  \citep[for reviews see][]{Fabian1994rev, Fabian2012, McNamara2012}. 
In this paper we examine the spectroscopic mass cooling rate in 3 well-studied, X-ray bright galaxy clusters, here abbreviated to Centaurus (A3526 $z=0.0104$), Perseus (A426 $z=0.018$) and A1835 ($z=0.235$).  

The 1999 launches of the Chandra and XMM-Newton Observatories gave arcsecond imaging  at CCD spectral resolution with ACIS on Chandra and high spectral resolution on an arcminute scale  with the Reflection Grating Spectrometers \citep[RGS;][]{denHerder2001} on XMM. It was soon realised that there was little direct evidence from either imaging or spectroscopy for complete cooling flows in cool cores, i.e of gas cooling down to 0.1 keV and  below. The basic picture of a cooling flow worked for the outer part of the cool core but the inner part with gas cooling below temperatures of 1--2 keV was at least complex. Large bubbles of jetted plasma, first seen with ROSAT in the Perseus cluster \citep{Boehringer1993} were found with Chandra to be common in cool cores  \citep[e.g.][]{Fabian2000, Mcnamara2000, McNamara2001, Blanton2001}. AGN (Active Galactic Nucleus) Feedback appeared to be responsible with the power implied by the bubbling matching the cooling luminosity \citep{churazov2002, Birzan2004}. Indeed bubbles were seen in all galaxy clusters and groups where heating was required \citep{Panagoulia2014}, apart from a very few objects such as A2029. The situation seemed to be understood apart from some details, which we examine here.

The first is how the bubbling actually heats the bulk of the gas in the cool core, which we addressed in 2017 \cite{Fabian2017}. 
The second is how such a tight balance between heating and cooling, with so little gas apparently cooling below 1 keV, has been established in the innermost cooler regions. The coolest X-ray emitting gas in clusters has cooling times down to below 10 million years. In particular the cooling time, and entropy, as a function of radius in a cool core has a similar shaped profile in most clusters irrespective of cluster mass \citep{Panagoulia2014, Babyk2018}. There is no sign of different states or flattening of the profiles, just similarity. Straightforward interpretation of RGS spectroscopy shows an inward  temperature drop from the higher bulk temperature down to 1--2 keV  but with any mass cooling rate diminishing  close to zero as the gas temperature drops further to 0.1 keV \citep{Peterson2001, Sanders2008}. \cite{Liu2019} found that the spectrally observed mass cooling rate below 1 keV in most of a sample of cool cores was less than 10 per cent of the values expected if there is no heating. There are nevertheless large quantities of cold molecular gas around the BCGs in cool cores with total cold gas masses ranging from $10^8$ to nearly $10^{11}\Msun$ \citep{Edge2001, Russell2019, Olivares2019}. Such gas is usually considered to be the result of cooling from the hot ICM. 

There are also significant masses, ($10^6 - 10^8\Msun),$ of both warm molecular and atomic gas detectable through emission lines in the optical, UV and infrared bands \citep[][and references therein]{Liu2020}. The regions around BCGs are highly multiphase. The low ionization state of the optical emission regions means that the excitation is not due to star formation \citep{Crawford1992} and may be due to energetic particles \citep{Ferland2009} or attenuated X-rays \citep{Polles2021}. Star formation is seen in some BCGs \citep{Johnstone1987, Mittal2015},  but it is certainly not ubiquitous. 

One possibility that we have considered is that the hot X-ray emitting gas cooling below 1 keV is physically mixing into the cold gas, thereby energizing it \citep{fabian2011,Fabian2012} . The energy that would have been emitted as soft X-rays is released as the long wavelength emission spectrum. It is however difficult to model and thus test such a process. In particular, there is as no clear evidence of charge exchange in the spectra  \citep[see e.g.][]{Walker2015,  Gu18}.

Here we investigate whether the apparent lack of soft-X-ray cooling gas, or "missing" soft X-ray emission, is due to photo-electric absorption, in an intrinsic process  where the emitting gas is closely interleaved with neighbouring cold gas. The intrinsic absorption model was shown by   \cite{Allen1997} to best represent ROSAT colour spectra of cool cores. \cite{Allen2000} used it successfully to model CCD resolution ASCA spectra. Simple additional absorption in cluster cores was first considered by \cite{White1991} for Solid State Spectrometer (SSS) spectra and by \cite{Fabian1994} for ASCA.  Intrinsic absorption was mentioned in a limited manner in several early RGS papers \citep[e.g.][]{Peterson2001} and briefly included in the multi-core study of \citep{Peterson2003} with most examples limited to columns of $N_{\rm H}<2\times 10^{20}\psqcm$.  A review by \cite{Peterson2006} notes that "no detailed test has been made of a model in which the absorption is {\em intimately linked} with the coolest cooling gas clouds." 

It is that model  we now explore here, using high spectral resolution RGS data. It is envisaged that regions within which  gas is radiatively-cooling below 1 keV and in which cold gas, including both that which cooled earlier and  stellar mass loss from the BCG, absorb the radiation produced. Mild turbulence can overlap different phases so that most soft X-ray emission regions are surrounded by cold regions giving intrinsic absorption. The idea has been successfully  applied to a filament in the Virgo cluster \citep{Werner2013} and more recently by \cite{Liu2021} to  the RGS spectrum of the massive cluster RXJC1504.1-0248 revealing a possible hidden mass cooling rate of $520\pm 30\Msunpyr$.   Here we apply the intrinsic absorption model to the 3 X-ray bright clusters mentioned earlier. 
 
\section{Spectral Analysis}

We use the XMM-RGS spectra of Centaurus, Perseus and A1835  reduced and analysed by JSS, see \cite{Sanders2008, Sanders2010} for details of the data reduction. Our aim is to see whether an intrinsic multilayer absorption model can allow significant mass cooling rates down to 0.1 keV. The presumption being that once the gas cools below the X-ray emitting band it will continue to cool to  $10^4$~K and below. 

The target clusters  have been selected to cover a range of star formation and AGN activity. Centaurus is a low mass cluster with a very low star formation rate (SFR) and a weak AGN, Perseus has a high  mass and SFR as well as a strong AGN; A1835 has a very high mass, high SFR  and a relatively weak nucleus. We defer exploration of a larger sample to later work. 

The intrinsic multilayer absorption model used by AF97 assumes emission and absorption regions are interleaved. Assume that each layer of emission at wavelength $\lambda$ contributes flux $\Delta F$ and each layer of absorption of column density $\Delta N$ transmits fraction $f=e^{-\sigma\Delta N}$ where $\sigma$ is the appropriate photoelectric absorption cross-section at $\lambda$, then we observe a total flux from $n$  layers 
\begin{equation}
F_{\rm t}=\Delta F+f\Delta F+f^2\Delta F+ ,,, f^n\Delta F.
\end{equation}
This geometric series reduces in the limit of large $n=N_{\rm H}/\Delta N$ to
\begin{equation}
F_{\rm t}=\frac{F_{\rm e}(1-e^{-\sigma N_{\rm H}})}{\sigma N_{\rm H}}.
\end{equation}
$F_{\rm e}=n\Delta F$ is the total flux emitted and $F_{\rm t}$ is the total flux observed. For spectral fitting with XSPEC it can be used as a multiplicative model using {\sc mdefine}\footnote{ mdefine mlayer (1-ztbabs(nh,z))/(-ln(ztbabs(nh,z))) :mul}.
Example spectra are shown in Fig. 1.

\begin{figure}
    \centering
    \includegraphics[width=0.54\textwidth]{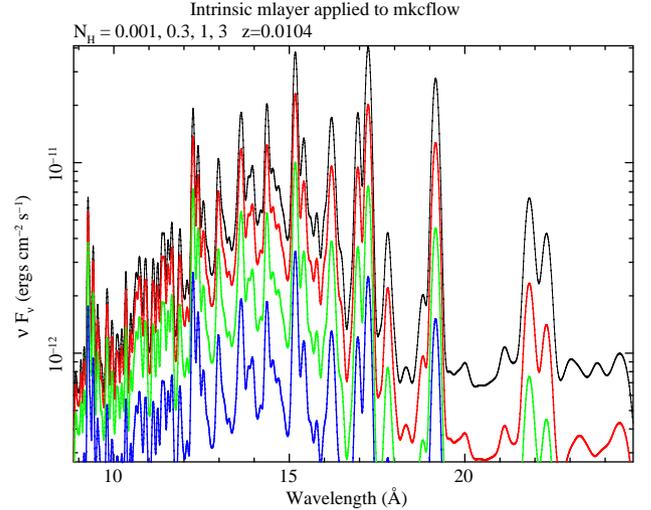}
    \caption{Multilayer, intrinsic absorption model applied to a cooling flow spectrum. Abundances are Solar. The total column density $N_{\rm T}$ goes from 0.001 (top) through 0.3 and 1 to $3\times 10^{22}\psqcm$ bottom). The strongest lines in the top spectrum are the 15\AA\ (resonance) and 17\AA\ lines of FeXVII which peak around temperatures of 0.5 keV and the 19\AA\ resonance line of OVIII which has a broader temperature distribution. }  
    \label{fig:my_label}
\end{figure}

\begin{figure}
    \centering
    \includegraphics[width=0.45\textwidth]{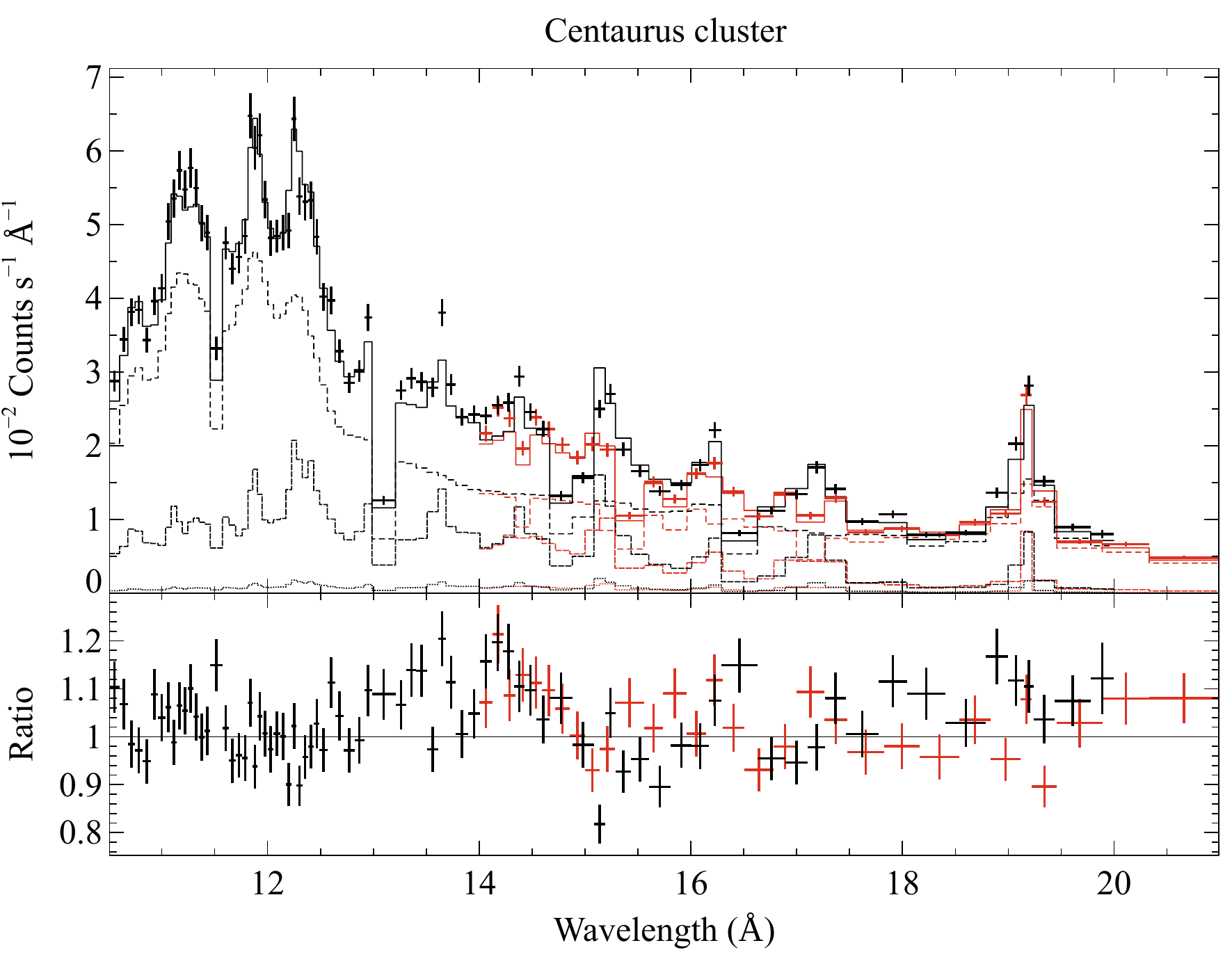}
        \includegraphics[width=0.45\textwidth]{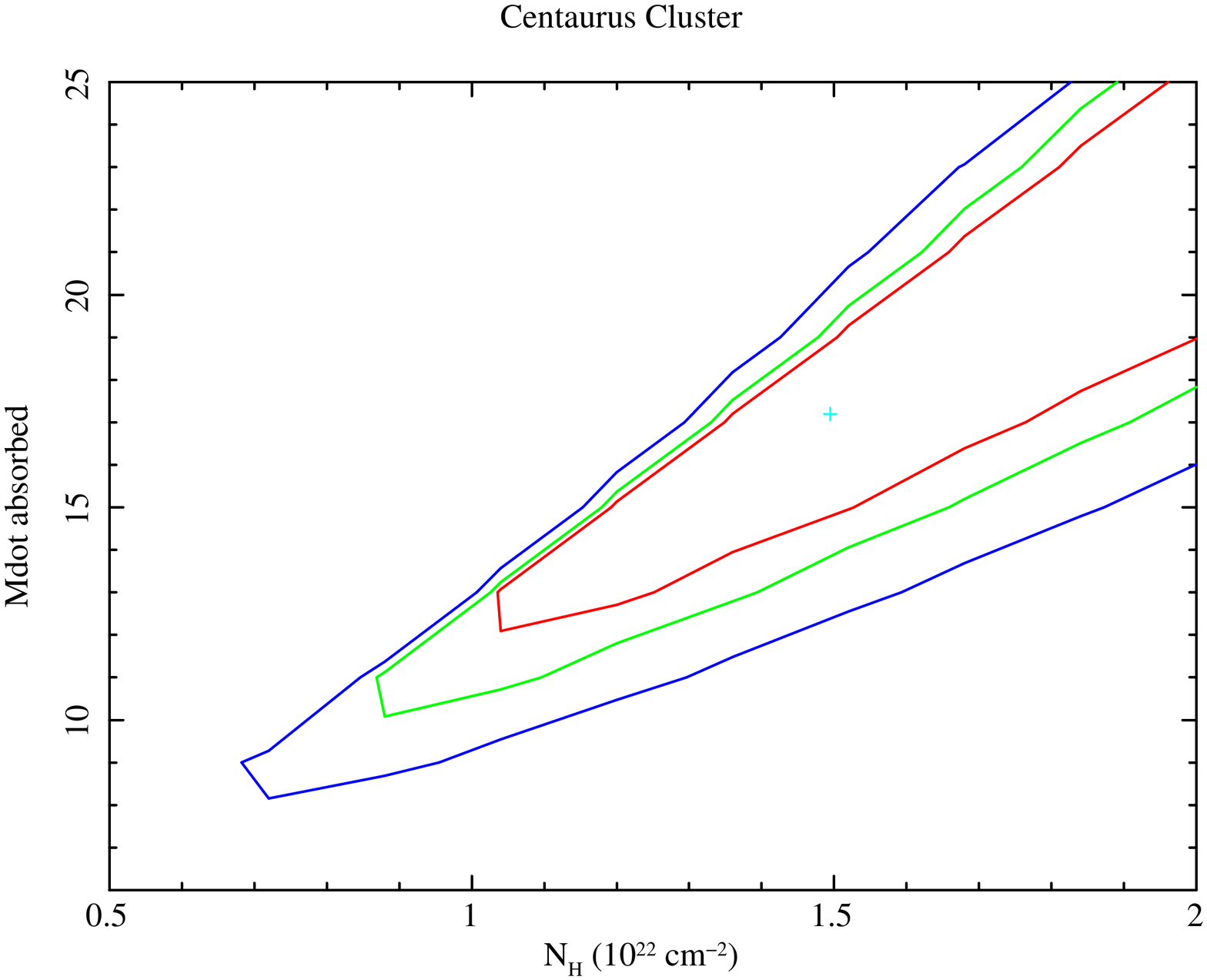}
     \includegraphics[width=0.45\textwidth]{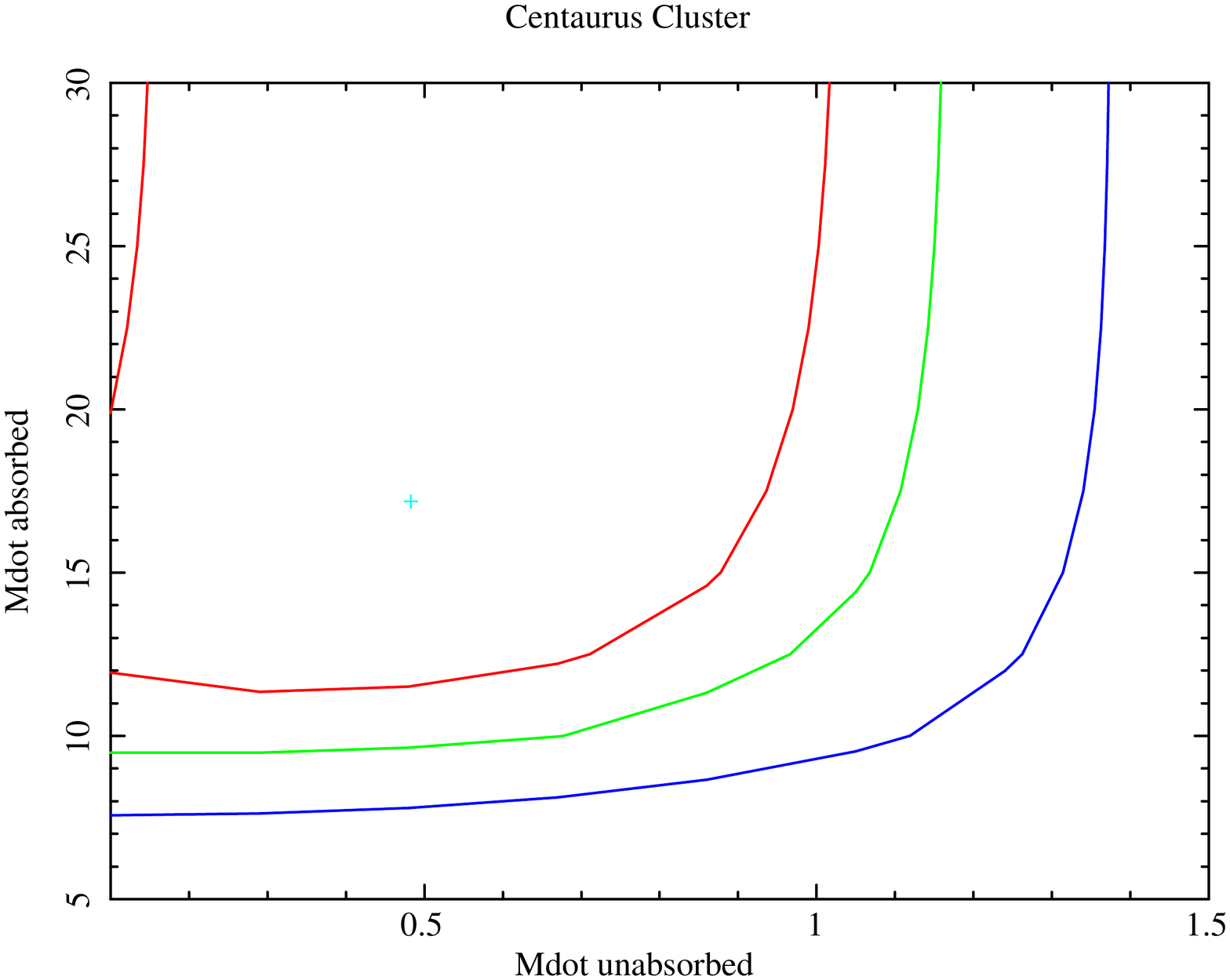}
    \caption{Top: Mean RGS 1 and 2 spectra of Centaurus \citep{Sanders2008}. The constant temperature, cooling and intrinsically absorbed components are shown as dotted lines. Middle: Absorbed mass cooling rate ($\dot M_{\rm a}$ in $\Msunpyr$) plotted against total intrinsic column density ($N_{\rm H}$ in units of $10^{22}\pcmsq$). Bottom: Absorbed mass cooling rate ($\dot M_{\rm a}$ in $\Msunpyr$) plotted against unabsorbed cooling rate ($\dot M_{\rm u}$ in units of $\Msunpyr$).
    }
    \label{fig:}
\end{figure}

Some investigation of absorption was carried out by \cite{Sanders2008} in which external absorption was applied to the 2 lowest temperature phases studied, 0.8--0.4 keV and 0.4--0.1 keV. This resulted in the allowable range of $\dot M$ for the first  phase being being about 4 times higher than the second. We apply the multilayer model to the 1 to 0.1 keV phase only. We can envisage a range of more complex models but this challenges the quality of the data. 

The RGS provides the spectrum of a slice across the complex centre of a nearby cool core. In the cross-dispersion direction, the aperture at 90, 95 and 99 percent, corresponds to approximately 0.8 arcmin, 1.7 arcmin and 3.4 arcmin wide strips. RGS 1 and 2 each have an (electronically) dead chip covering 10.4--13.8 A and 20--24 A, respectively. 

Where absorption is interspersed with soft X-ray emission there will be some emission viewed unabsorbed and some from the other side viewed through considerable absorption. The ratio of emission to absorption may be a function of depth, as may be the temperature of the emission. The intrinsic absorption model is just a first approximation of this complexity \footnote{We note that the emission from  mixing layers \citep{Begelman1990} may also be intrinsically absorbed.}. Our goal here is to test the simplest model and obtain limits on the total mass cooling rates.  

The model we adopt and apply to all 3 clusters has 
3 components consisting of a single  temperature thermal gas, a cooling flow from that temperature to 1 keV and an intrinsically absorbed cooling flow of the same mass cooling rate ($\dot M$) cooling from the single temperature to 0.1 keV. In {\sc XSPEC} terms this is {\sc gsmooth*apec + gsmooth*(mkcflow +mlayer*mkcflow)}. We  add Galactic absorption and also separate Gaussian smoothing ({\sc gsmooth}) of the hotter and cooler components  to compensate for the effect of the spatial extent of the source (the RGS is a dispersive spectrometer) and any intrinsic turbulent blurring.  In figures we show the spectrum, the absorbed mass cooling rate $\dot M_{\rm a}$ plotted against the intrinsic absorption column density $N_{\rm H}$ and $\dot M_{\rm a}$ plotted against the unabsorbed mass cooling rate $\dot M_{\rm u}$. The cooling rates are mostly limits contoured at 68, 90 and 99 per cent confidence levels.

For Centaurus we use data from the widest 99 per cent aperture and match parameters with those in the earlier analysis \citep{Sanders2008}. Results for the cooling rates from the smaller 90 per cent aperture are similar (most of the coolest gas lies close to the centre). The quality of the fit (i.e. reduced  $\chisq$ value) is not very good at 1.26, but similar to that of the earlier analyses \citep{Sanders2008}. 
The spectral fit and cooling rate contours are shown in Fig. 2.  The absorbed cooling rate is $\dot M_{\rm a} = 14.2\Msunpyr$ if the intrinsic $N_{\rm H}=1.2\times 10^{22}\psqcm$.

Perseus is more distant and we use the 95 per cent aperture spectrum. It is complicated by having an X-ray bright AGN. However XMM has pn CCD detectors as well so the AGN contribution can be well characterised at CCD resolution. \cite{Reynolds2021}  found that a partial-covering model is appropriate for the Perseus nucleus using XMM and Chandra grating spectra (covering fraction of 0.15 with $N_{\rm H}=5.7 \times 10^{23}\psqcm$). We use their result here for the 2001 data. We use the Hitomi temperature for the outer gas (3.5 keV {\sc apec} component).   The spectrum and cooling rate contours are shown in Fig. 3. The absorbed cooling rate $\dot M_{\rm a}= 35.6\Msunpyr$ (if $N_{\rm H}=2.5\times 10^{22}\psqcm $) and   the unabsorbed rate is  then $\dot M_{\rm u}<6.7\Msunpyr.$

\begin{figure}
    \centering
    \includegraphics[width=0.45\textwidth]{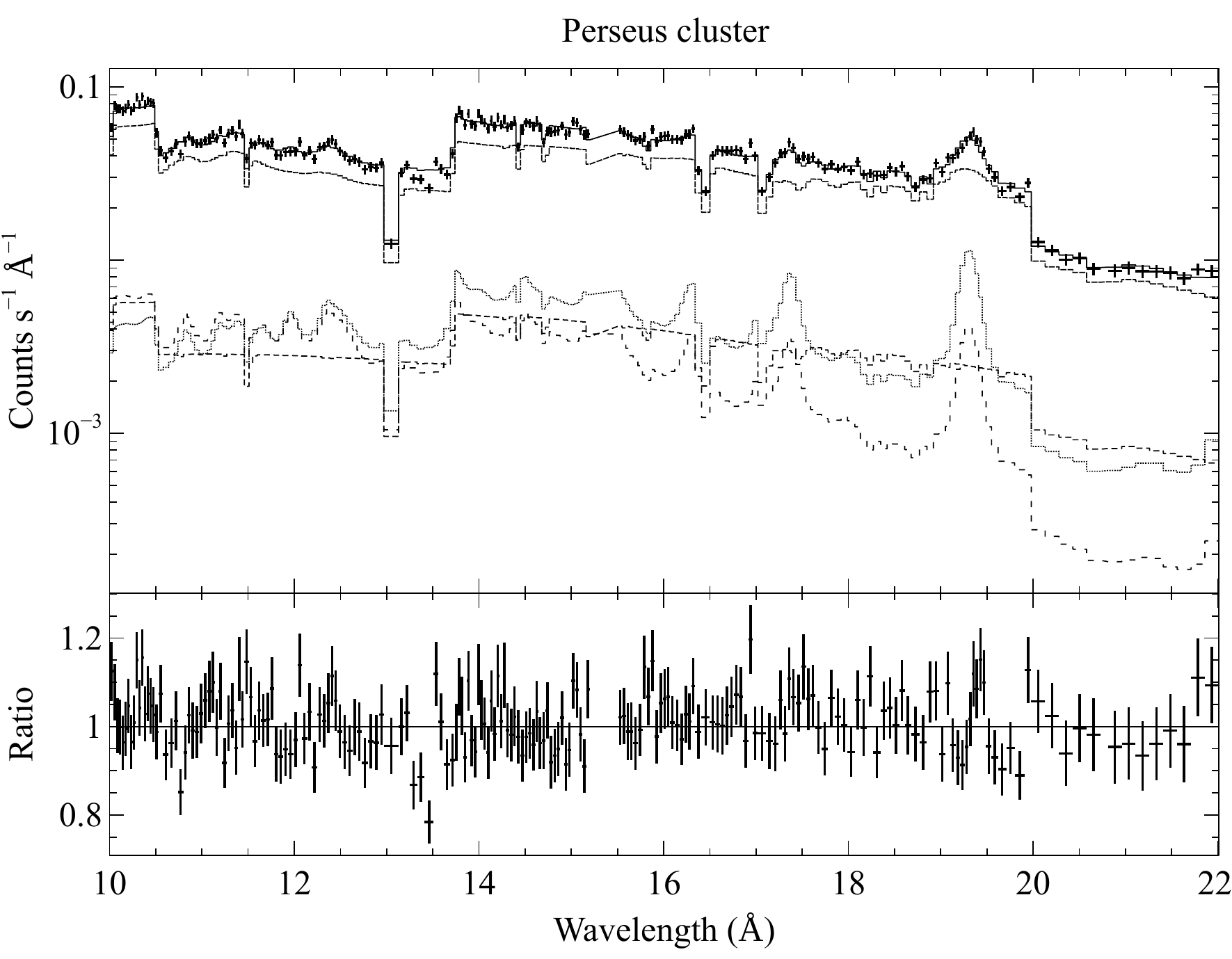}
        \includegraphics[width=0.45\textwidth]{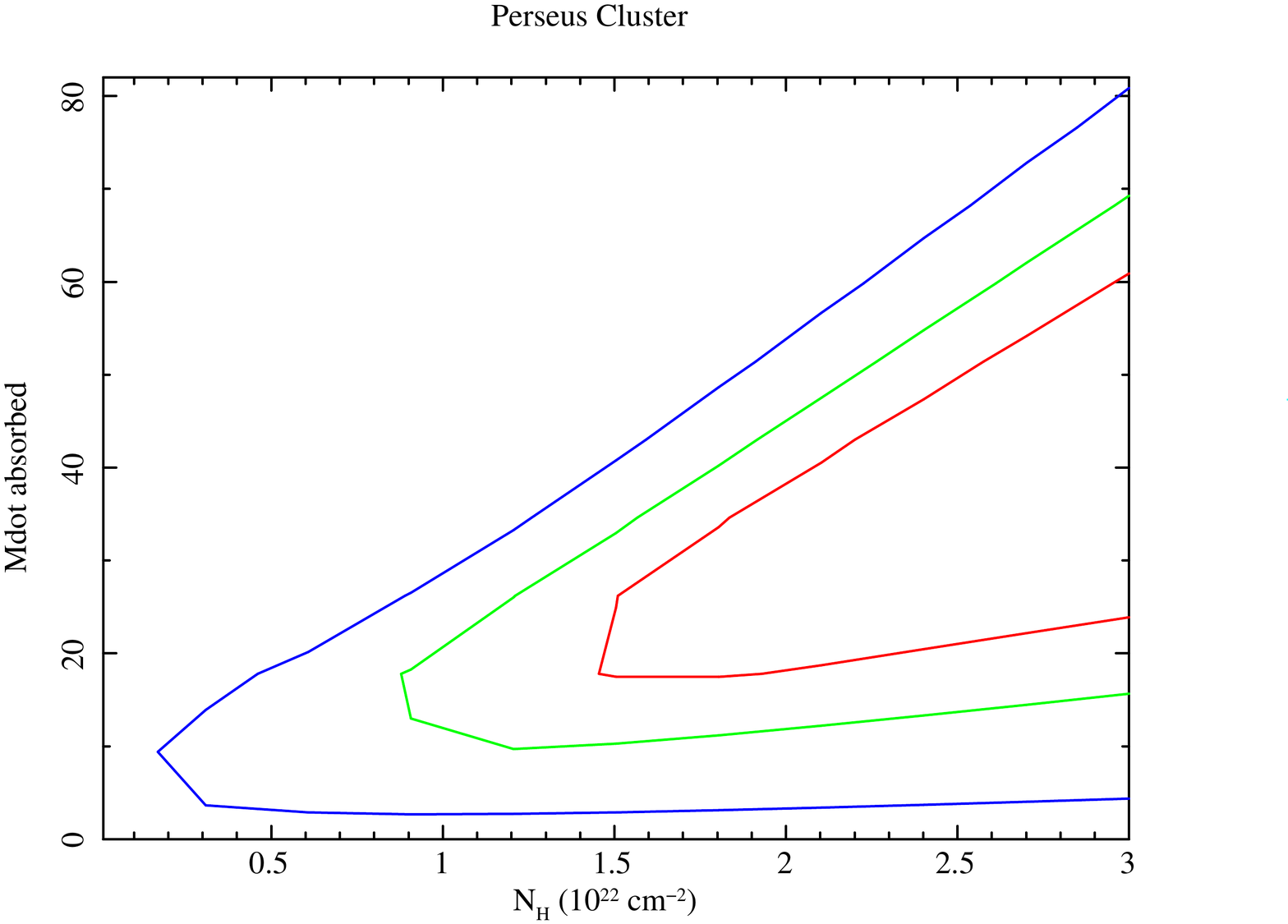}
     \includegraphics[width=0.45\textwidth]{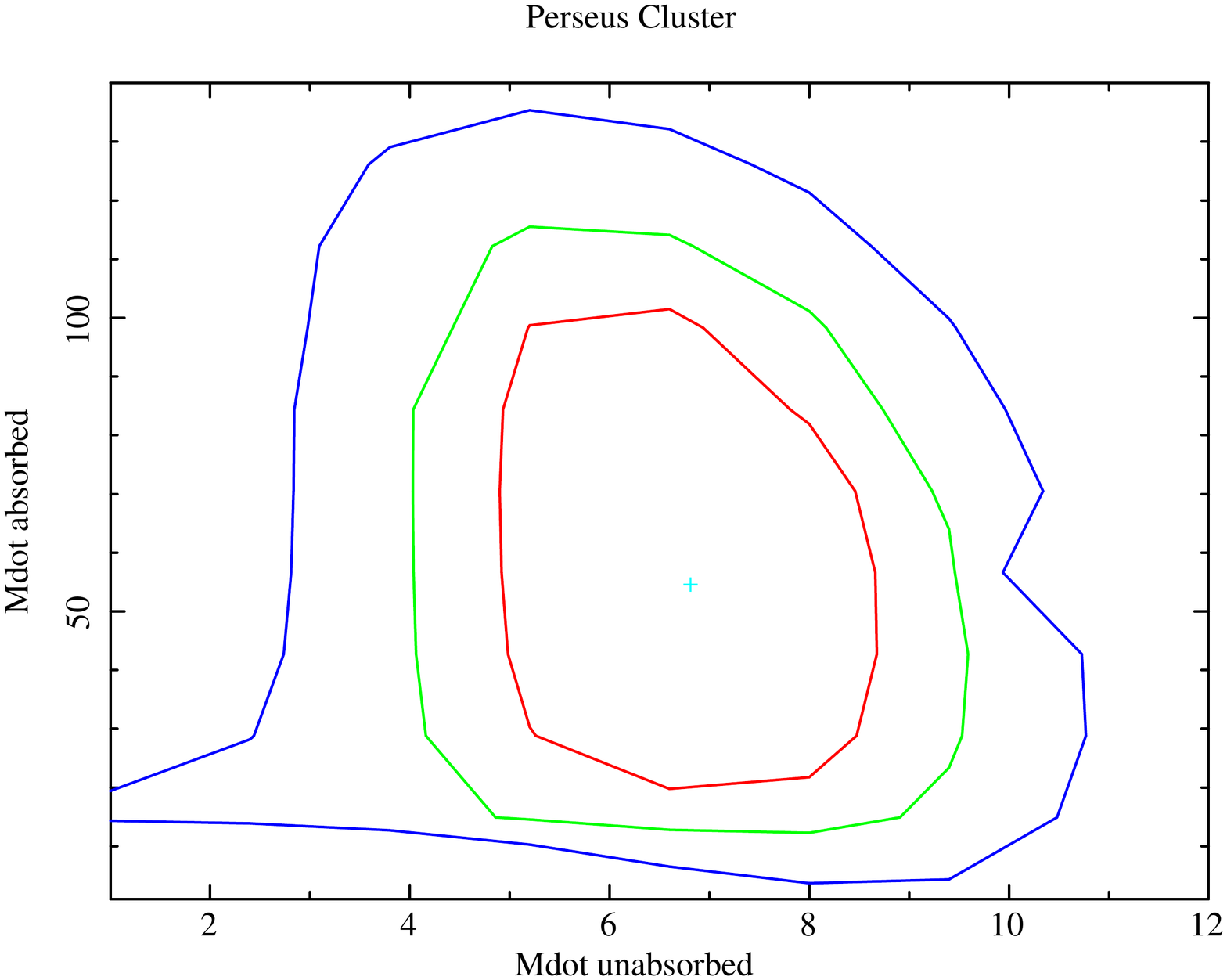}
    \caption{Top: Combined RGS 1 and 2 spectra of Perseus. The constant temperature, cooling and intrinsically absorbed components are shown as dotted lines. Middle: Absorbed mass cooling rate ($\dot M_{\rm a}$ in $\Msunpyr$) plotted against total intrinsic column density ($N_{\rm H}$ in units of $10^{22}\pcmsq$). Bottom: Absorbed mass cooling rate ($\dot M_{\rm a}$ in $\Msunpyr$) plotted against unabsorbed cooling rate ($\dot M_{\rm u}$ in units of $\Msunpyr$).
   }
    \label{fig:my_label}
\end{figure}

Emission lines at 15 and 17\AA\ due to FeXVII are prominent  in the model spectrum. The resonance one at 15\AA\ (15.2\AA\ in our frame) leads to a negative dip in the residuals. The observed high forbidden (17\AA) to resonance (15\AA) line ratio has been noted by \cite{Pinto2016} and explained by resonance scattering\footnote{Optical depth effects in the hot component become important if its column density exceeds $10^{22}\pcmsq$ \citep{Chakraborty22}.}. The added presence of intrinsic absorption means that the scattered line will be rapidly absorbed, rather than escape.  We excise a small region around the 15A line to avoid that model line dominating the fits. The poor fit around the base of the 19\AA\  OVIII line is due to the degeneracy in blurring caused by spatial and spectral line broadening in a dispersive spectrometer. A variable abundance cooling flow model ({\sc vmkcflow}) was used here as the oxygen abundance appears to be higher then that of iron.  

\begin{figure}
    \centering
    \includegraphics[width=0.45\textwidth]{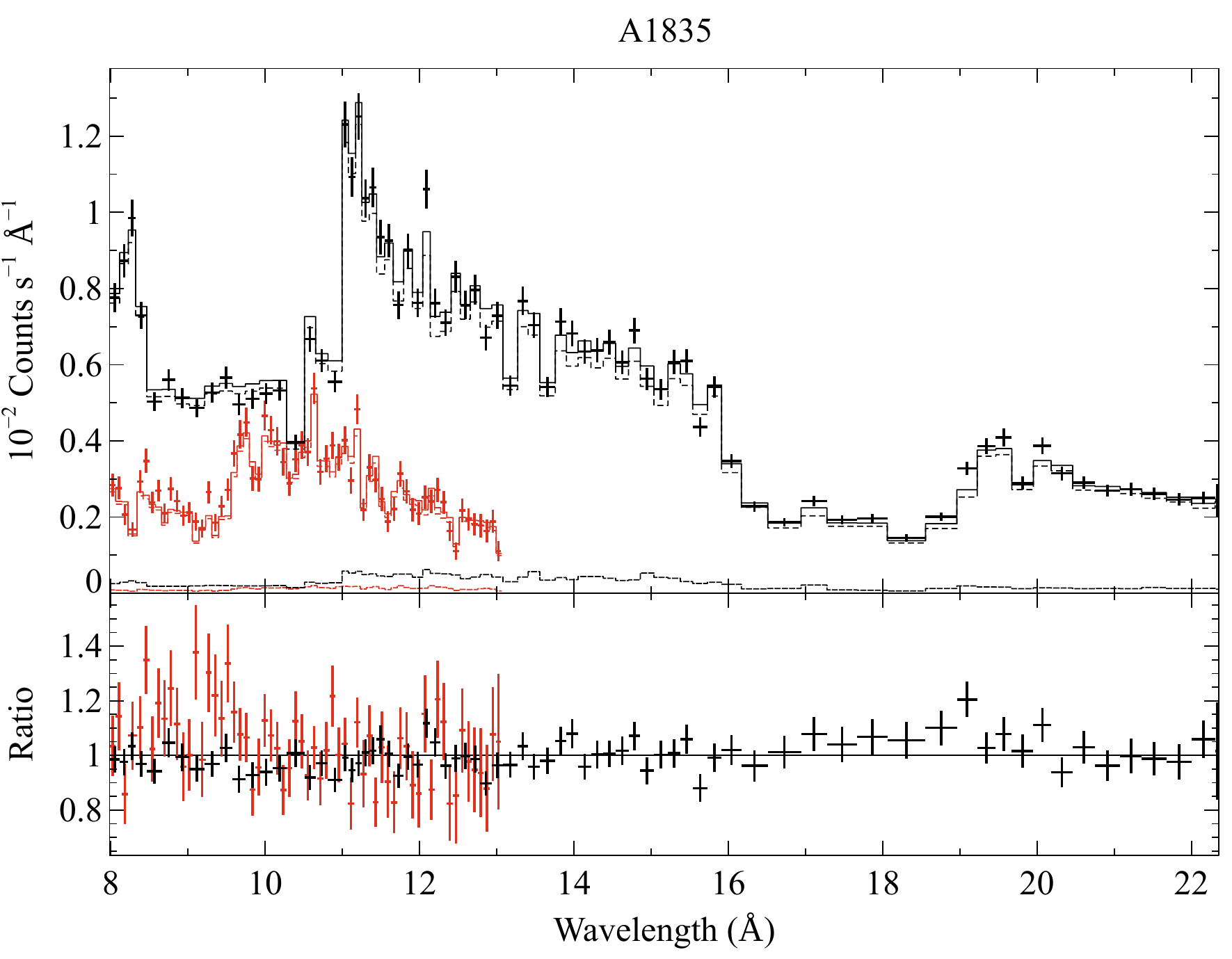}
        \includegraphics[width=0.45\textwidth]{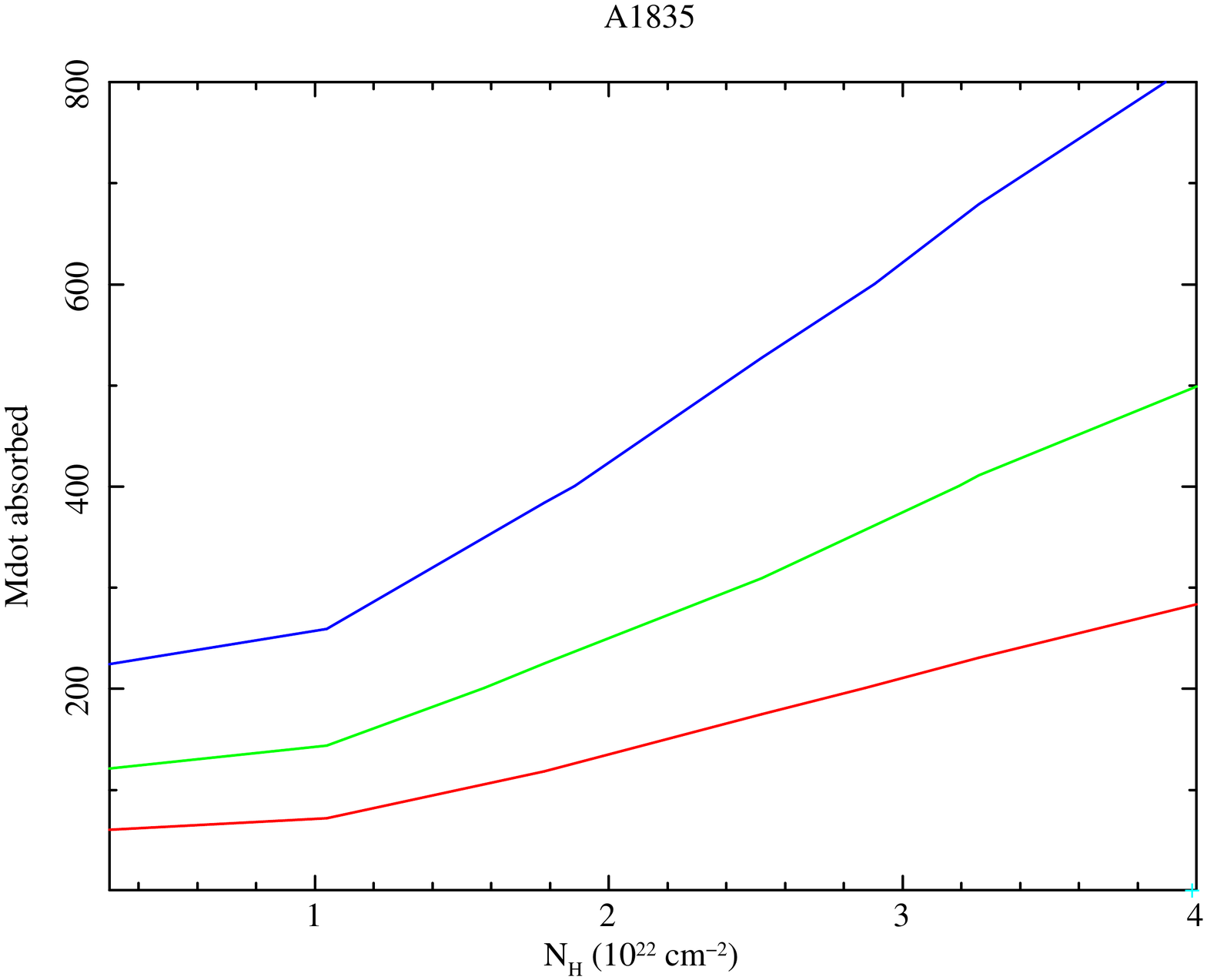}
    \includegraphics[width=0.45\textwidth]{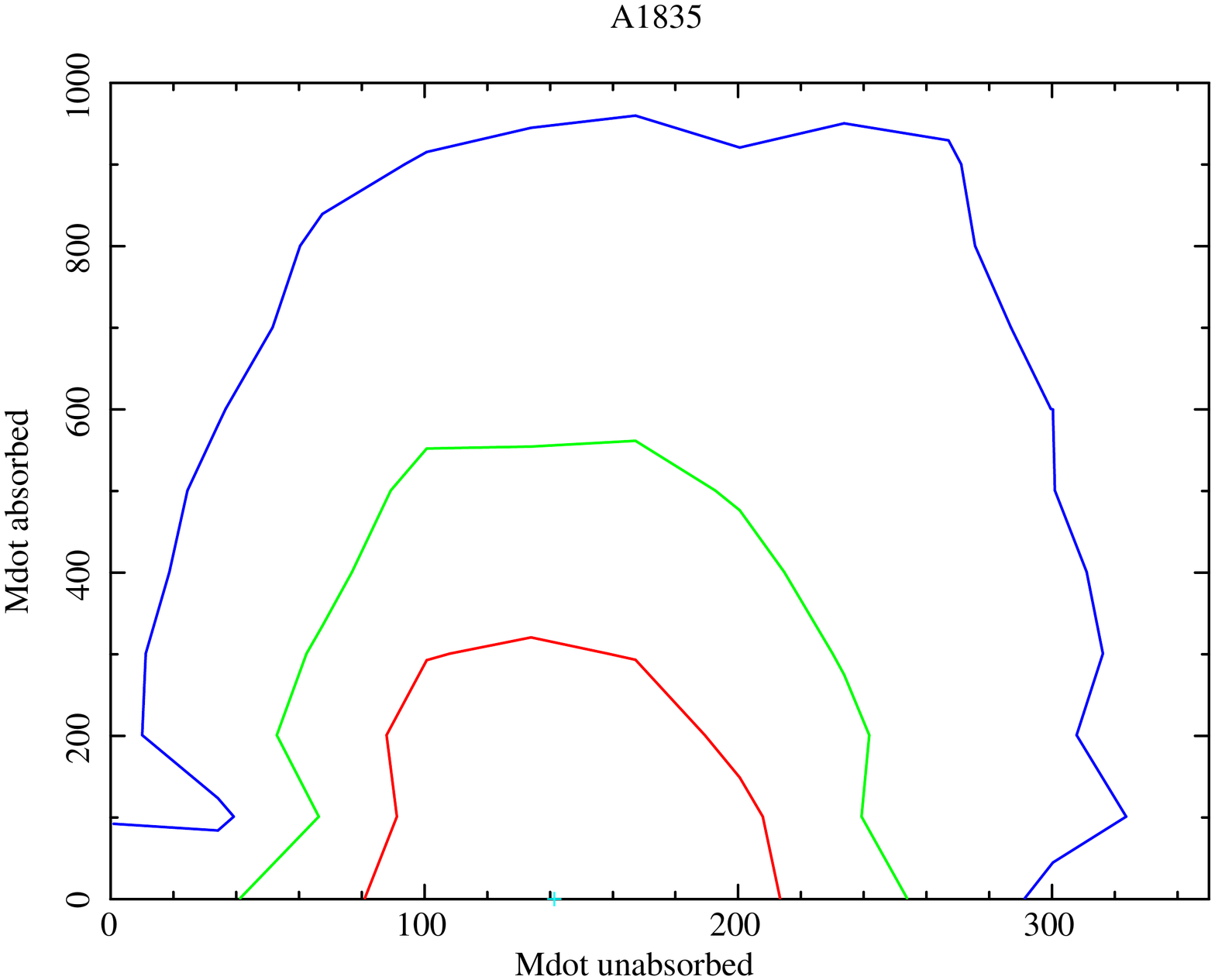}
    \caption{ Top: Mean RGS first and second order spectra of A1835 \citep{Sanders2010} plotted against rest wavelength (just for this object). The constant temperature, cooling and intrinsically absorbed components are shown as dotted lines. Middle: Absorbed mass cooling rate ($\dot M_{\rm a}$ in $\Msunpyr$) plotted against total intrinsic column density ($N_{\rm H}$ in units of $10^{22}\pcmsq$). Bottom: Absorbed mass cooling rate ($\dot M_{\rm a}$ in $\Msunpyr$) plotted against unabsorbed cooling rate ($\dot M_{\rm u}$ in units of $\Msunpyr$).}
\end{figure}

For distant A1835 we use the same approach with the 90 per cent aperture.  There is no bright nucleus and we take values from the earlier fits on the same data by \cite{Sanders2010}. We use Cash statistics here since the flux is low. This time there is a significant unabsorbed component that in the fitting trades off with the intrinsically-absorbed one (Table 1 and Fig 4.). If the intrinsic absorption is fixed at $N_{\rm H}=4\times 10^{22}\psqcm$ the unabsorbed rate  $\dot M_{\rm u}=142\pm86\Msunpyr$ and the absorbed value of  $\dot M_{\rm a}<330\Msunpyr,$ bringing the total up to $\dot M \sim 500 \Msunpyr.$
 
\begin{table}
	\centering
	\caption{Spectral Fitting Results. The units of column density are $10^{22}\psqcm$, $kT$ is keV, $Z$ is Solar abundance and $\dot M$ is $\Msunpyr$. The smoothing kernels, Sigma$_1$ and Sigma$_2$, are defined in keV at 6 keV. The redshifts were fixed (f).  } 
	\label{tab:properties_table}
	\begin{tabular}{lccr} 
		\hline
		Parameter & Centaurus & Perseus & A1835  \\
		\hline
		z & 0.0104f &0.01076f & 0.253f \\
		Galactic $N_{\rm H}$ & 0.12f &0.12f&0.02f\\
		 Sigma$_1$ & $(5.6\pm0.5)$e-2 &$(7\pm1.5)$e-2& 1.74e-2f\\
		Apec kT & $1.70\pm0.03$& 3.5f& $3.76\pm0.23$\\
		$Z_{\rm O}$ & $0.43\pm0.02$ & $0.5\pm0.06$ & 0.2f\\
			$Z_{\rm Fe}$ &" & $0.32\pm0.04$ & "\\
		Apec Norm &$(1.55\pm0.02)$e-2&$(6\pm0.2)$e-2&$(1\pm0.03)$e-2\\
		 Sigma$_2$ &$(1.2\pm0.05)$e-2 & 2.54e-2& 1.75\\
		 	$\dot M_{\rm u}$ & $0.5\pm0.4$ &$6.7\pm2.2 $ & $142\pm86$\\
		 		 	$\dot M_{\rm a}$ & $14.2\pm 2$ & $35.6\pm18$& $<330$\\
		Intrinsic $N_{\rm H}$ &1.2f&2.5f & 4f\\
		$\chi^2/{\rm dof}$ & 1457/1168 &1227/1147 & \\
		${\rm C}^2/{\rm dof}$ & & & 2596/2571\\
		\hline
	\end{tabular}
\end{table}

\subsection{The outer filaments in Perseus}

\cite{Walker2015} have summed Chandra spectra of the outer filaments in Perseus and show that they can be fitted by a two temperature model, with temperatures of 0.8 and $1.95 \keV$. We have fitted this spectrum with {\sc apec + mlayer*mkcflow} and obtain constraints on absorbed cooling with $\dot{M} \sim 1.5\Msunpyr$ and intrinsic $N_{\rm H}\sim 3.5 \times 10^{21}\psqcm$ (Fig.5). This is a minor contribution to the overall solution obtained from the RGS spectra.

\section{Comparison with FIR luminosities}

We find that the multilayer absorption model  allows significant mass cooling rates to occur below $\sim 1$ keV, much of it hidden from direct view. The question then arises as to how large  the rate can be. Clearly energy has to be conserved and that is where the total emission from longer wavelengths can provide the constraint. The  regions we are observing are dusty and  Far InfraRed (FIR) emission is very important. All 3 clusters have been observed with the Spitzer and/or Herschel telescopes and we list their FIR luminosities in Table~2. We include the corresponding mass cooling rate from 1 keV using equation 1, $\dot M_{\rm FIR}$, using  1 keV which is appropriate for the column densities considered.

In Centaurus we see striking agreement, if most of the FIR luminosity is due to absorption of the cooling X-ray emission, rather than star formation and the intrinsic column density $N_{\rm H}< 1.5\times 10^{22}\psqcm$.  \cite{Mittal2011} deduced a SFR of $0.13\Msunpyr$ if it accounts for all the FIR emission and much lower values of $0.002-0.08\Msunpyr$ from the observed FUV emission. The range of the latter estimate allows for a dominant cooling flow contribution.
There is no detectable X-ray emission from the nucleus which is also weak in the optical band.

In Perseus the FIR luminosity \citep{Mittal2012} exceeds the cooling luminosity by a large factor  if $\dot M<100 \Msunpyr$. However the BCG, NGC1275, has an A-star spectrum and there is, and has been, much ongoing star formation \citep{Mittal2015}. Also there is a luminous AGN at the centre (NGC1275 is an original Seyfert galaxy). 

In A1835 we find  $\dot M\sim300\Msunpyr$, possibly approaching $800\Msunpyr$ if intrinsic absorption is included.
The FIR luminosity \citep{Egami2006} far exceeds any due to cooling.

\begin{figure}
    \centering
    \includegraphics[width=0.45\textwidth]{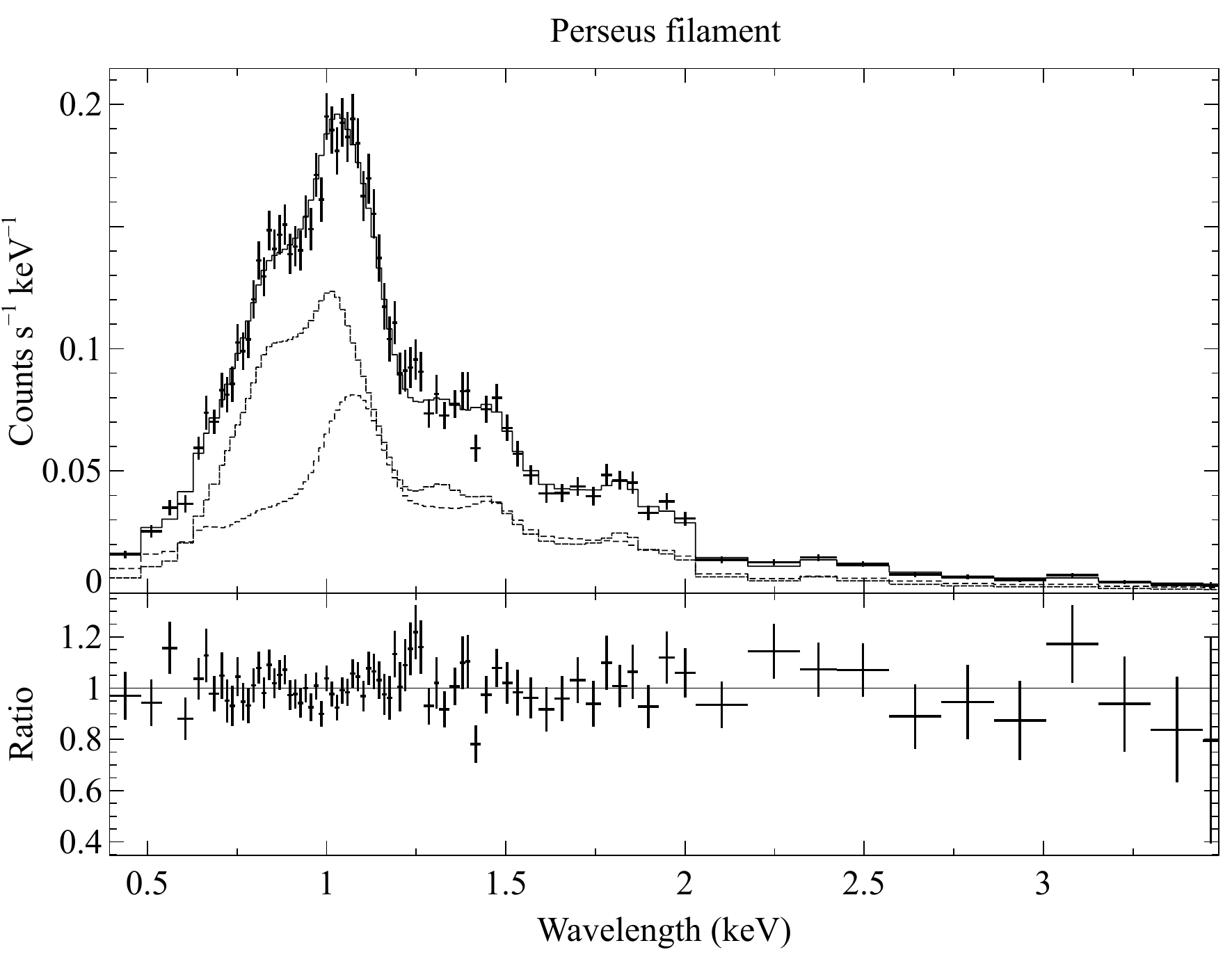}
\includegraphics[width=0.45\textwidth]{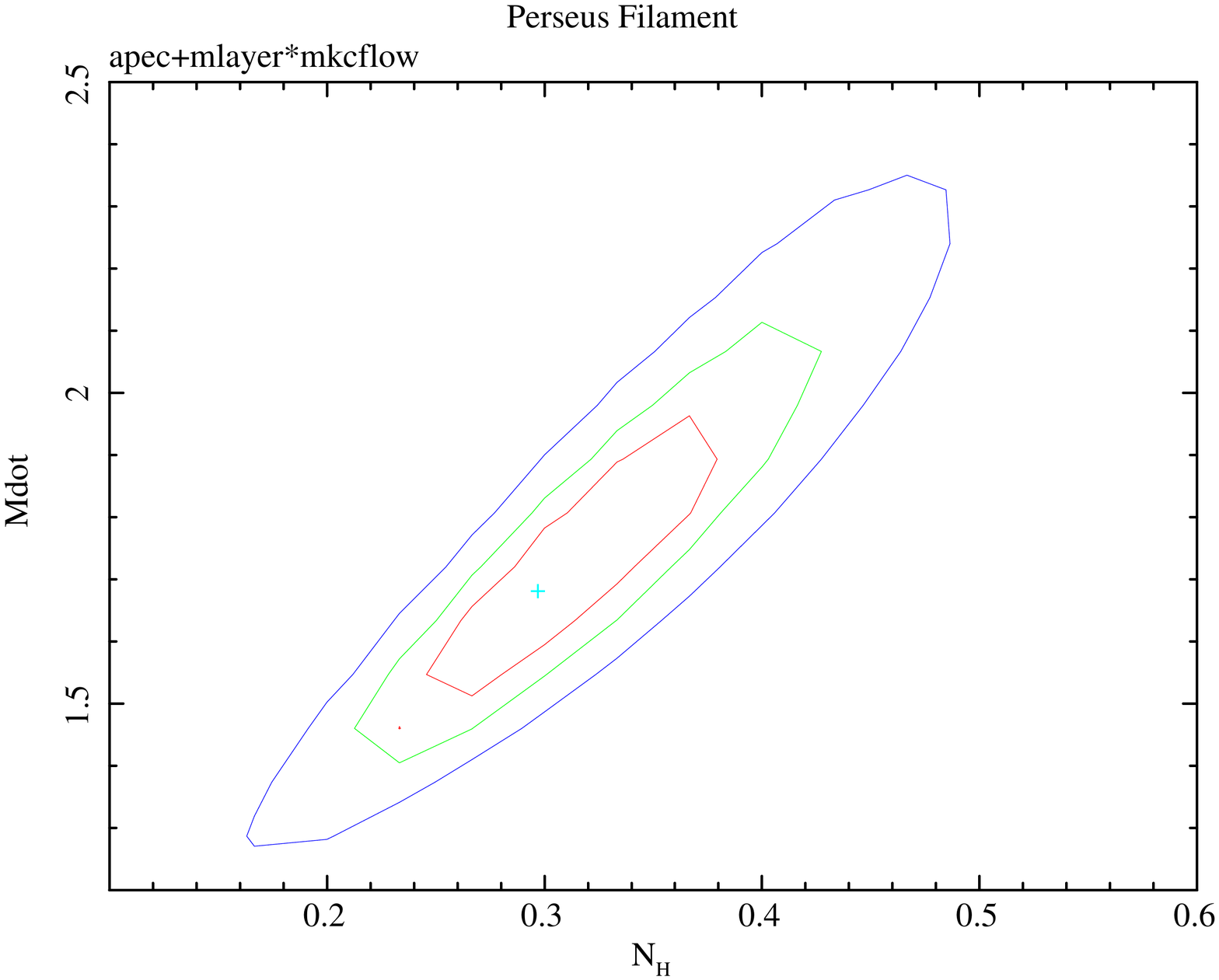}
    \caption{Top: Summed Chandra spectrum of selected outer filaments in Perseus, as shown in \citep{Walker2015}. The constant temperature and absorbed cooling components are shown as dotted lines.   Bottom: Allowed regions for the absorbed  $\dot{M}$ and intrinsic column density $N_{\rm H}$ within the filaments. }
    \label{fig:my_label}
\end{figure}

\section{Implications}
The mass cooling rates we have found imply the accumulation of  large cold masses if they continue for several Gyr. Star formation is an obvious endpoint for some of the gas, and is relevant for Perseus and A1835.   For Centaurus the simple implication is of a large accumulated cold mass of $1.5\times 10^{10}\Msun$ per Gyr. The molecular mass detected in Centaurus by ALMA and reported by \cite{Olivares2019} \citep[see also APEX result in ][]{Fabian2016} is only $10^8\Msun$, which means that over 100 times that mass is missing unless it is either hidden from view and/or there is  a destruction mechanism as discussed below, or it changes state (collapses into brown dwarfs or planets). We discuss all possibilities below.

In Centaurus there is a dusty filamentary optical structure of atomic gas \citep[][and references therein]{Fabian2016} and  a recently discovered faint {\em diffuse} H$\alpha$ and N[II] nebula discovered in  MUSE data by \cite{Hamer2019}. The diffuse gas extends to the North beyond the filamentary structure and  presumably consists of clouds with  a molecular core which would be undetected if very cold and undisturbed. 

In Perseus, the  $2\times 10^{10}\Msun$ of CO-detected molecular gas already detected by \cite{Salome2006} would be replenished in  a Gyr, which is much less than the age of the cool core. There may again be a significant unseen mass of cold gas. Possible evidence for its existence lies in the apparently isothermal shock seen around the inner bubbles \citep{Fabian2006, Graham2008}. Although a density jump of about 30 per cent is seen, no significant temperature jump is measured. This would result if there is a cold gas fraction of about 20 per cent by mass, forced to mix into the hotter gas by the passage of a shock. Tentative evidence for this explanation is seen by the abrupt ending of the Northern filaments at the shock front, as discussed by \cite{Graham2008} and shown in Fig. 6. 

Further unusual filament behaviour in Perseus is revealed at the shock front in the West and South (see Fig 6). Here the initially radial filaments fan out tangentially over several kpc at the shock. The velocity of these regions determined from optical spectroscopy is unexceptional \citep{Gendron-Marsolais2020}. The only plausible explanation for such behaviour is that the cold gas was already present before the shock passage and has been rendered detectable by the passage of  the shock front which has excited it. A Galactic analogy is the Western Veil Nebula of the Cygnus Loop where the shock front has overtaken cool interstellar clouds \citep{Fesen2018}. The atomic gas mass responsible for the appearance of these features is small and we assume that all such gas is accompanied by much more molecular gas.

This leads to the conclusion that the total mass of cold gas is {\em reduced}  by the shocks produced by the bubbling process as an integral part of AGN feedback. Some of the hot gas cools from above $10^7$~K down to $30$~K, and is later heated back up to above $10^7$~K by mixing following the passing of shocks.

The high pressure region behind the isothermal shock front (shown in Fig. 6) already contains more than three-quarters  of the energy of the bubbles \citep{Graham2008}. As the shock propagates outward it will weaken and essentially become an adiabatic sound wave incapable of destroying cold clouds. If it can dissipate it will heat the general cool-core ICM \citep{Fabian2017}.

\begin{table*}
	\centering
	\caption{Relevant Cluster Properties. See \citep{Mittal2011, Mittal2012, Mittal2015, Olivares2019} for Centaurus and Perseus, \citep{Egami2006, McNamara2006}for A1835. The (small) uncertainties in the values for Perseus and A1835 are dominated by modelling systematics.}
	\label{tab:properties_table}
	\begin{tabular}{lcccccr} 
		\hline
		Cluster & L(FIR) &$\dot M_{\rm FIR}$ & Dust Mass & CO Mass & SFR \\
		\hline
		 & $\Lsun$ &$\Msunpyr$& $\Msun$ & $\Msun$ & $\Msunpyr$\\
		Centaurus &$(7.9\pm2.5) {\rm e}8$& $12.8\pm4$ & $(1.8\pm0.3){\rm e}6$ & 1e8 & <0.13\\
		Perseus &1.5e11& 2.1e3 & 1.e7 & 2e10 & 20-40 \\
		A1835 & 8e11 &1.1e4 & 1e8 & 5e10 & 100-180\\
		\hline
	\end{tabular}
\end{table*}

\begin{figure*}
    \centering
    \includegraphics[width=0.95\textwidth]{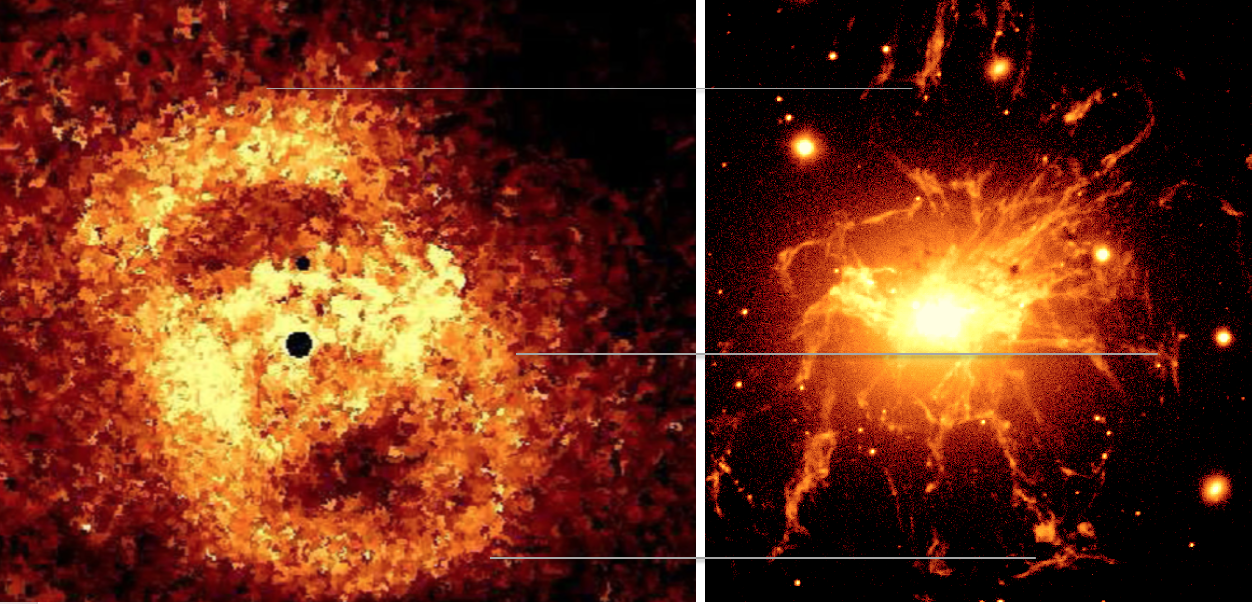}
%    {\includegraphics[width=0.45\textwidth]{PerN.png}
%    {\includegraphics[width=0.45\textwidth]{Perw.png}
%            \vspace{0.1cm}}
%    \includegraphics[width=0.45\textwidth]{PerS.png}
    \caption{Left: Pressure map for Perseus core from \citep{Fabian2017}, the outer edge of the two thick, bright rings is the shock front. Right: HST H$\alpha$ +N[II] emission nebulosity centred on NGC1275 \citep{Fabian2008}, the Perseus BCG. Thin horizontal lines indicate from top, the end points of the Northern filaments which stop at the shock; middle: The W tangential filaments and lower: the S tangential filament. The images are $46\kpc$ from top to bottom. }
    \label{fig:my_label}
\end{figure*}

\subsection{Comparison with the Crab Nebula}
The low ionization spectrum characteristic of the filamentary nebula in cluster cool cores, e.g. with strong low ionization line ratios such as NI/H$\alpha$,  is unlike that of HII regions, and is more similar  to that of the  Crab Nebula. Both lie in regions dominated by a central object emitting nonthermal particles, a pulsar in the case of the Crab and a jetted AGN in the case of cluster cool cores. The environment in both has aspects of a pulsar wind nebula. \citet{Ferland2009} proposed that the excitation of cool-core filaments is due to energetic particles. Emission from the cold molecular knots in the Crab may have a similar origin \citep{Richardson2013}. These knots have now been strongly detected in CO with ALMA \citep{Wootton2022} and, surprisingly, are at a similar thermal pressure ($ n T\sim 10^{6.5}\pcm$~K) to those in  the Perseus cluster (and other cool core clusters). 

One implication of such sub-Solar mass cold clouds is that they exist and can survive in the most surprising environment. We do not know the detailed mass structure of the filaments and clouds in cluster cores. Hubble Space Telescope (HST) observations resolve the optical filaments to have widths of about 75~pc in Perseus and 50~pc in Centaurus \citep[][respectively]{Fabian2008, Fabian2016}. They cannot however be of uniform composition or they would be impossibly massive and instead  consist of many much smaller threads and knots. Magnetic fields must play a role in maintaining their integrity although the details of that how that works are unclear. 

Another implication is from the dust associated with many of the Crab knots. Significant amounts of dust must have formed in them over the past 1000 yr. If it forms rapidly there then why not in cold dense clouds around BCGs?
Further study of these relatively nearby examples of cold dense  molecular clouds could help to define their mass spectrum, detailed structure, origin and fate. 

\subsection{Very cold clouds}
In Fig. 7 we show the cooling rate, cooling time and emission fractions for cold clouds at the pressure expected near the centre of a cool core ($ n T\sim 10^{6.5}\pcm$~K). The cooling time below 30~K is short at about 30~yr. Unless disturbed or irradiated, such clouds should cool quickly to undetectability. We suggest that such clouds may be common in the inner regions around BCGs of cool core clusters.

We have in the past studied the fate of such cold clouds, whether dusty \citep{Fabian1994dust} or dust-free \citep{Ferland1994, Ferland2002}. The total mass of such clouds could be tens of per cent of the hot gas mass, which in the case of Perseus  is $5\times 10^{10}\Msun$ within 10 kpc. As explored by \cite{Ferland1994}, the Jeans mass for clouds at the temperature of the Cosmic Microwave Background temperature and pressure assumed here is less than $0.1 (1+z)^2\Msun$. Very cold clouds that collapse and fragment will rapidly form low-mass stars and substellar objects such as brown dwarfs and planets. See \cite{Jura1977, Fabian1982, Sarazin1983} for early discussion of star formation under these conditions and \cite{Sharda2022} for a recent study.

Evidence for a "bottom-heavy" Initial Mass Function (IMF) in early-type galaxies was obtained by \cite{vDC2010} and later work; the situation has been reviewed by \cite{Smith2020}. \cite{Oldham2018} find radial gradients in the stellar mass-to-light ratio, $\Upsilon$, in M87, the nearest BCG. Their modelling has $3.5\times 10^{10}\Msun$ of stars within the inner kpc where $\Upsilon$ is highest.

\section{Discussion}

We have shown that significant mass cooling rates may be occurring in cool core clusters, provided that the gas cooling below about 1 keV is interspersed with cold absorbing clouds having total column densities of $\sim10^{22}\psqcm$. This leads to the accumulation of much larger masses of cold gas than observed or generally considered so far. The total mass of cold gas is controlled by a combination of cloud destruction by shocks in the AGN bubbling process and collapse of the coldest clouds into brown dwarfs and planets. This picture can account for the apparent isothermal shock in the Perseus cluster and the diffuse H$\alpha$+N[II] gas seen beyond the main filamentary nebula in Centaurus. 

There is little evidence for shocks around the inner bubbles in most clusters, possibly because most observations are not deep enough or the cluster lies too far away for any bubble rims to be resolved. Where there are deep observations of nearby objects the situation is confusing. The Centaurus cluster shows bright, high pressure, rims around the inner bubbles \citep{Sanders2016}, but deprojected spectra of the clearest part to the NE shows a gradual temperature rise only downstream of the density jump but not at the jump itself. In A2052, \cite{Blanton2011} find two density jumps with the inner one showing a temperature rise at a  significance of $2.1\sigma$ and the outer one having a brief temperature rise that is less than half the expected value. \cite{Randall2015} present results for the multiple bubble system in the galaxy group NGC5813. They find, at high significance, that the amplitude of the temperature jumps are below those expected for shocks in adiabatic gas. In contrast
M87 has a classical spherical shock with density and temperature jumps in accord with theoretical expectation  at a radius of 13 kpc \citep{Forman2007, Forman2017}. It lies well beyond the bubbling source or "piston" in that system. Any multiphasedness as envisaged here does not extend to such relatively large radii.  

\begin{figure}
    \centering
    \includegraphics[width=0.4\textwidth]{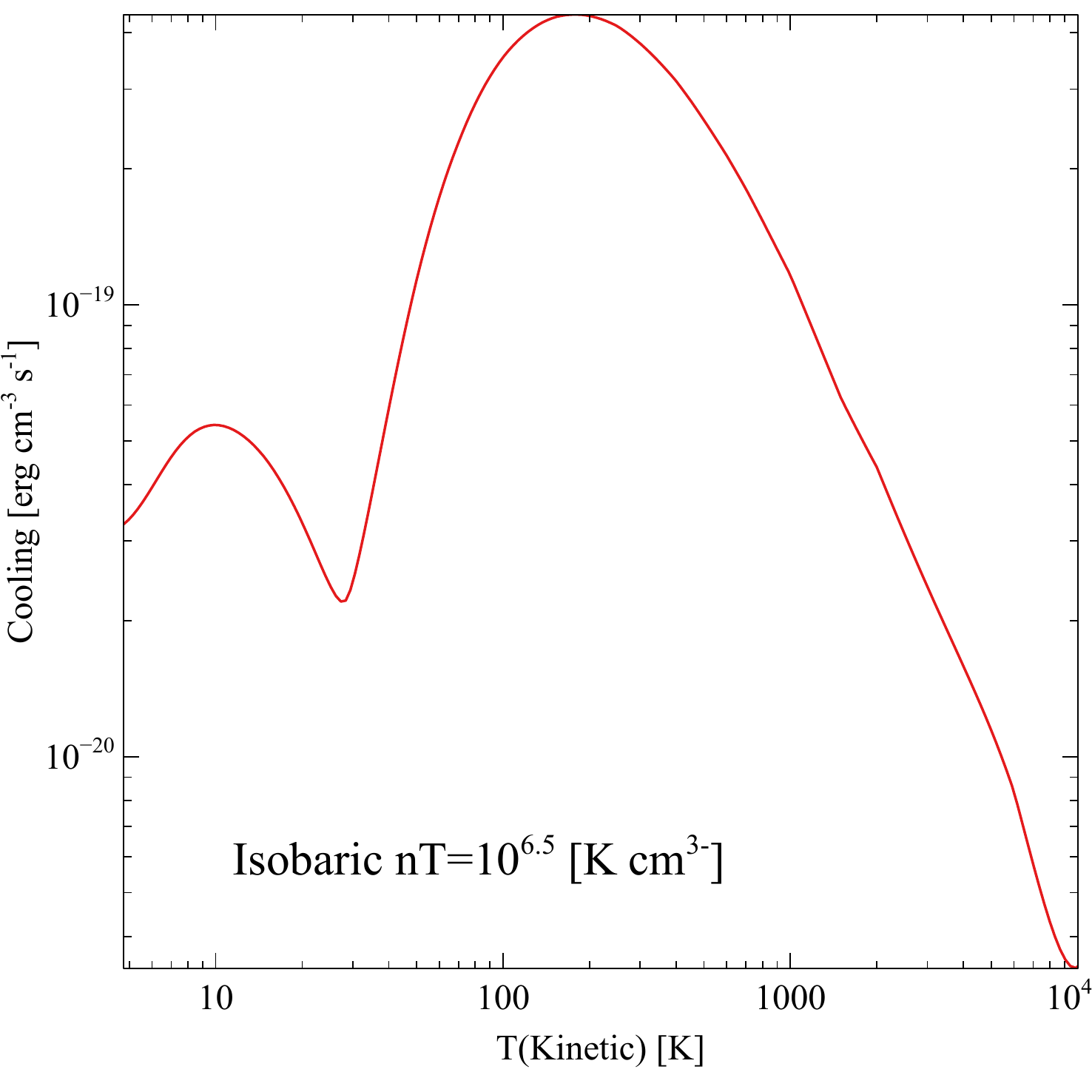}
      \includegraphics[width=0.4\textwidth]{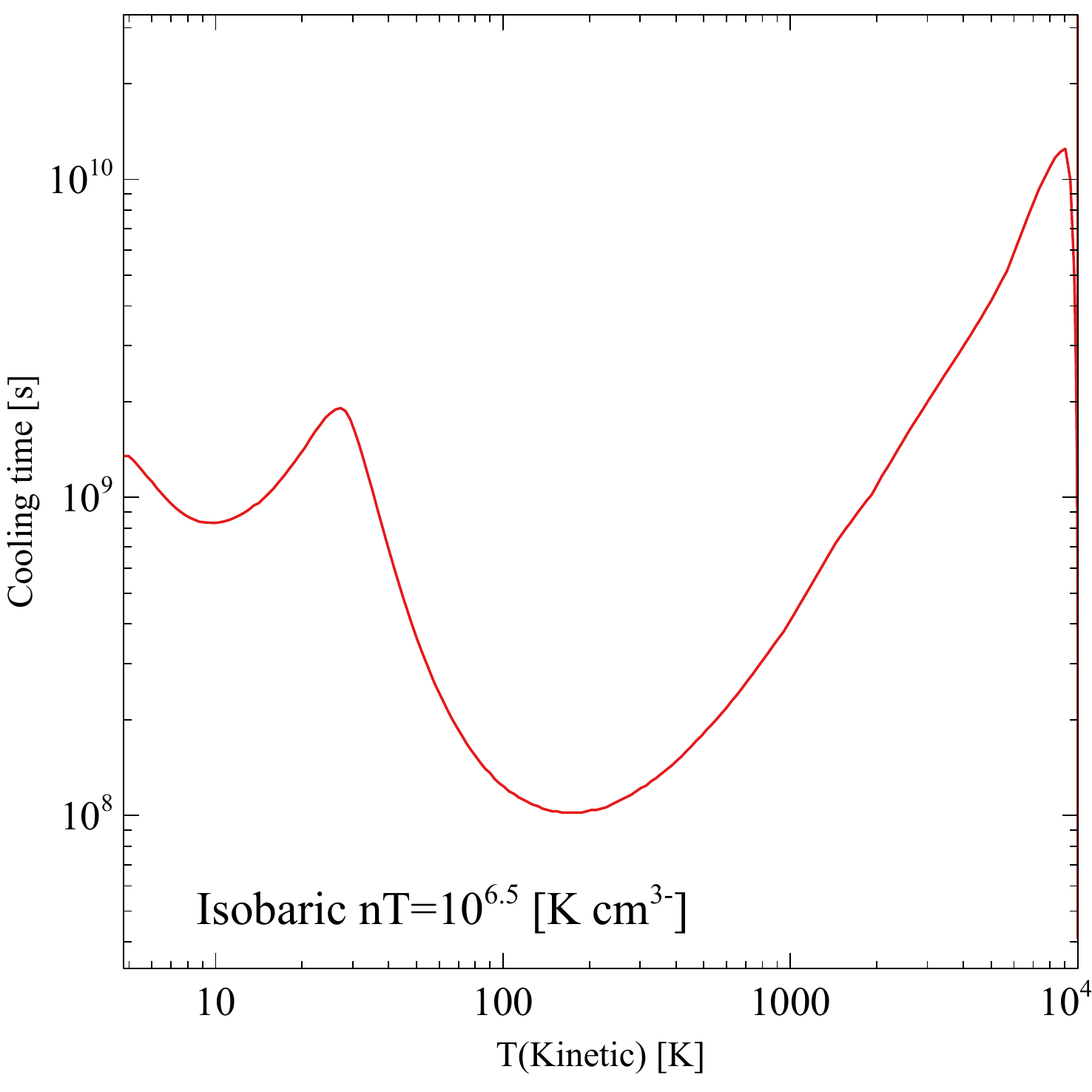}
     \includegraphics[width=0.4
    \textwidth]{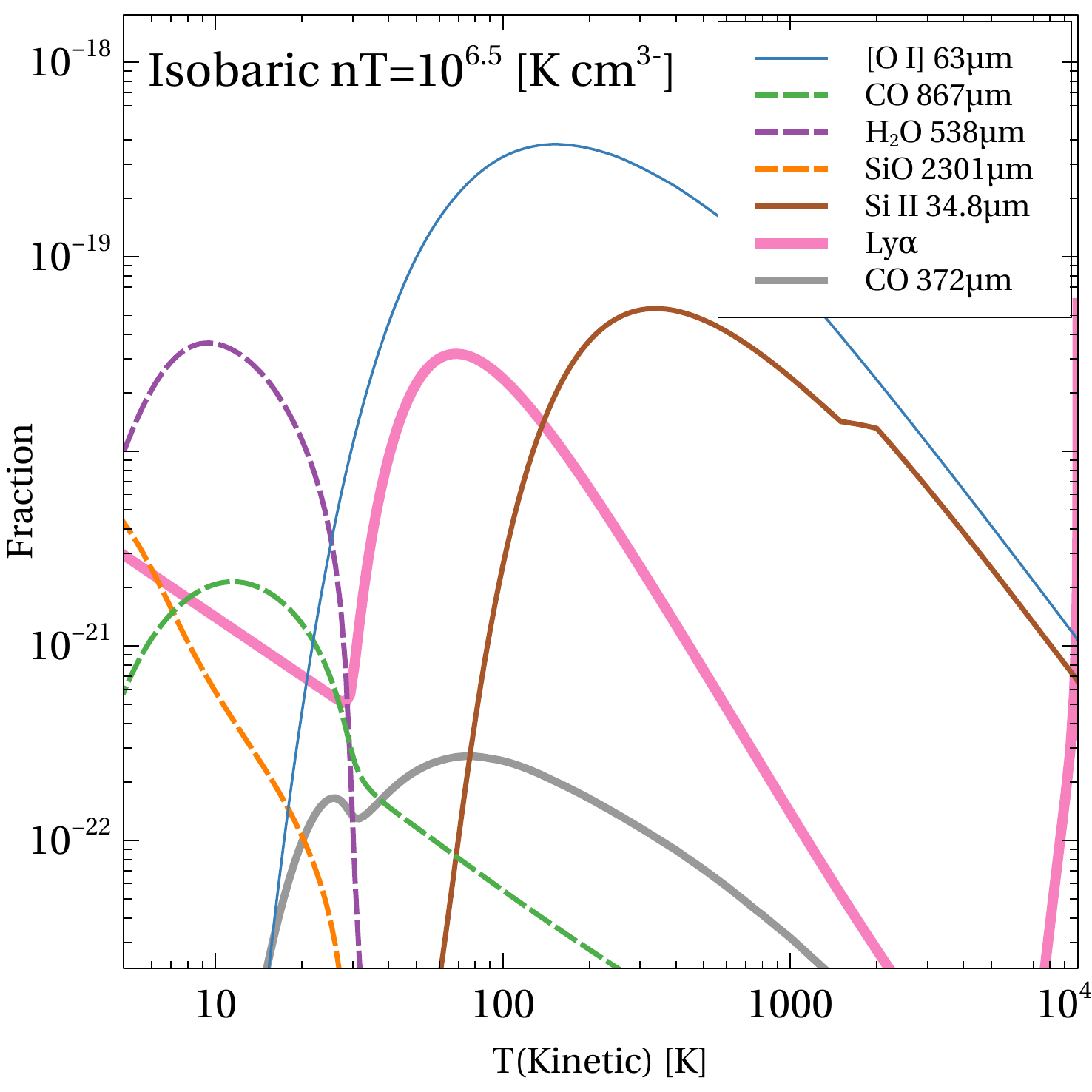} 
    \caption{Constant pressure cooling function and cooling time for cold gas at constant pressure $nT=10^{6.5}$. The lower panel shows the dominant emission lines.} 
    \label{fig:my_label}
\end{figure}

We propose that the hot phase in the inner few kpc of cool cores contains many undetected cold clouds. The enormous density contrast of nearly a million between the cold clouds below 50~K and the hot gas at $10^7\K$ means that the cold clouds are very small in size. They may be freely floating or  more likely are magnetically-coupled to each other or the hot gas. If and when they agglomerate into larger clouds and filaments may then become detectable. The larger observed clouds and filaments are where more normal star formation occurs.  Such agglomerations may be {\em transient}, dissolving and reforming, which may help to explain some of the their puzzling dynamics  \citep{Russell2019}. 

Cold clouds may also be seeded with dust ejected by the older stars in the BCG as well as dust formed within the clouds as in the Crab Nebula. This  explains why the gas is dusty and readily forms molecules. Persistent clumps of stellar mass loss may aggregate and act as dusty nuclei for cold clouds. Steeply rising abundance gradients are commonly seen in the hot gas in nearby cool cores, where they can be spatially resolved \citep{Panagoulia2015}. The gradients reveal a cycle in which rising bubbles drag cool dusty gas out from the centre until they mix with the hotter gas \citep{Panagoulia2013, Panagoulia2015, Lakhchaura2019, ALiu2019}. Such cold clouds would be unusually dust rich. Their atomic emission may also account for the diffuse emission seen by \citet{Hamer2019}. 

One indication of this is obtained by considering the gas to dust ratio of the absorbing matter. In Centaurus we are looking at the inner kpc where the Herschel PACS FIR 100$\mu m$ emission originates \citep{Mittal2011}. The mass of molecular gas within that central region from the \citet{Olivares2019} CO observations is about $4\times 10^7\Msun$.  The dust mass estimated by \citet{Mittal2011} is $2\times 10^6\Msun$, yielding a gas to dust ratio of 20 (a low value is also suggested by \citet{Mittal2011} on the basis of the C[II] emission).  This effect will reduce the hydrogen column density of the cold gas so that  the "effective" column density is $2\times10^{21}\psqcm$ instead of $10^{22}\psqcm$,

We estimate the mass of cold absorbing gas, $M_{\rm c}$ from  the volume, $V$,  of the hot cooling gas obtained by  considering just the thermal energy and equating
$$n^2V\Lambda= \frac{3}{2} \frac{\dot M kT}{\mu m}.$$
Taking $\Lambda=2\times 10^{-23}\ergcmcups$ \citep{Sutherland1993} for the cooling function at the appropriate metallicity (about half Solar) and  mid-temperature ($5\times 10^6\K$), $\dot M=\dot M_1 10\Msunpyr$ and $nT=10^{6.5}\pcmK$, we find that $V=\dot M_1 10^{65}\cmcu$. Assuming a sphere,  the radius is then
$\dot M_1 $ \kpc. The cold absorbing gas is assumed to be uniformly spread through that volume so for a column density of $N_{22}=10^{22}\pcmsq$, $M_{\rm c}=4\times 10^8 N_{22} {\dot M_{1}}^{2/3}\Msun.$ These values approach  the the size and observed mass of the central molecular region in Centaurus, especially if the effective column density is reduced as suggested in the previous paragraph.  

The Centaurus cluster is nearby which means that we have good spatial resolution of the components in its innermost regions. It also has a weak AGN and little star formation. In comparison, Perseus is complicated by the bright central AGN and star formation. The much greater distance to A1835 (more than an order of magnitude) coupled with strong star formation makes any detailed considerations difficult. Their gas-to-dust ratios, simply estimated from the FIR flux and CO detections, are >1000 \citep{Edge2001}, This does not necessarily apply however to the relatively small regions where any hidden cooling flow is occurring. Further observations are required.

A luminous and very active central AGN such as seen in the exceptional Phoenix cluster \citep{McDonald2019} can distrupt the cold gas distribution and reveal the soft X-ray cooling \citep{Pinto2018}. Interestingly, patches of X-ray absorption have been mapped in that cluster, coincident with the soft X-ray emission. Column densities of about $10^{22}\psqcm$ are measured for that absorbing gas \cite{McDonald2019}. \citet{Walker2015} also found  a similar column density for an extended region in  the Perseus cluster.  

Absorption of the central AGN in cool core clusters is expected to be common with hidden cooling flows. We have already mentioned the partial covering of the Perseus nucleus found by \cite{Reynolds2021}. Photoelectric absorption is one explanation for the lack of X-ray bright nuclei in many cool cores \citep{HL2011}. Molecular line absorption is now being increasingly detected through ALMA observations \citep{Rose2019MNRAS}. It is a powerful probe of the nature of the cold clouds.

Perhaps there  is no feedback cycle as such, but just a continuous jet from the central engine \citep{McNamara2011}. The intermittency of the bubbles is just due to bubbling (as in a fish tank aerator, or a dripping tap). There may of course be variations in the jet power but it  would always be on. The power may originate in the spin of the black hole and released via the Blandford-Znajek mechanism.

Major progress in the issues discussed here requires deeper observations of cool cores both in the X-ray band and at long wavelengths, e.g. with ALMA and perhaps JWST. The properties of the multiphase interstellar gas and of the stellar populations are important as especially is the mass-to-light ratio $\Upsilon$ of the central regions. The simple modelling carried out here can hopefully be applied to nondispersive, microcalorimater, high resolution X-ray spectra over the whole 0.3-10 keV band using XRISM \citep{XRISM2018} after its 2023 launch. Deeper Chandra observations of cool core clusters, including Perseus, are also needed to study the isothermal shocks. Cool cores will be excellent targets for future  missions like Athena \citep{Barcons2017} and concepts like AXIS \citep{AXIS2018}, 

Our results raise the strong possibility that there may be significant levels of hidden (absorbed) cooling. We find a total  $\sim15\Msunpyr$ cooling flow in Centaurus  and possible flows of $30-100\Msunpyr$ in Perseus and $\sim500\Msunpyr$ in A1835. Heating from AGN feedback reduces the overall mass cooling rate by a factor of 2--3, much less than the factor of 10 or more obtained with no hidden cooling.  We also introduce the possibility that within  the inner regions of cool cores there are significant additional masses of cold gas  that have so far evaded detection, due to their very low temperature and  cold and weak emission. The coldest clouds may collapse and fragment into low-mass stars and  substellar objects, contributing to the mass of dark matter already present in BCGs. We suggest that AGN feedback controls the overall level of cold gas through the bubbling process, which  shocks some clouds and drags others to larger radii. The "missing" soft X-ray emission emerges in the FIR to UV bands in these objects.  

\section{Conclusions}
Hidden (absorbed) cooling flows may occur in cool core clusters, alleviating the fine-tuning implied by the widespread apparent lack of gas cooling below 1 keV. AGN Feedback is still needed to prevent massive cooling rates in the hotter gas. The cooled gas accumulates in cold clouds which are either destroyed by the bubbling process caused by AGN Feedback, or collapse rapidly to form low mass stars and brown dwarfs.

\section*{Acknowledgements} We thank the referee for a helpful report. ACF thanks Keith Arnaud for help with {\sc MDEFINE}. BRM acknowledges the Natural Sciences and Engineering Research Council for their support.

\section{Data Availability} All data used here are available from ESA's XMM-Newton Science Archive and NASA's Chandra Data Archive.

\bibliographystyle{mnras}
\bibliography{cool_core_Mdot} % if your bibtex file is called example.bib

% Don't change these lines
\bsp	% typesetting comment
\label{lastpage}
\end{document}